\documentclass[a4paper,10pt,twoside]{cpc-hepnp}

\usepackage{multicol}
\usepackage{graphicx}
\usepackage{booktabs}
\usepackage{amssymb,bm,mathrsfs,bbm,amscd}
\usepackage[tbtags]{amsmath}
\usepackage{lastpage}

\begin{document}

\fancyhead[co]{\footnotesize V.D. Burkert:  N* physics at Jefferson Lab}


\title{The N* Physics Program at Jefferson Lab}
\author{ Volker D. Burkert   \\ (for the CLAS Collaboration) }
\maketitle
\vspace{0.2cm}
\address{Jefferson Laboratory \\12000 Jefferson Avenue, Newport News, VA23606 \\
\footnote{e-mail: burkert@jlab.org}}

\begin{abstract}
Recent measurements of nucleon resonance transition form factors with CLAS at Jefferson
 Lab are discussed.
The new data confirm the assertion of the symmetric constituent quark 
model of the Roper as the first radial
excitation of the nucleon. The data on high $Q^2$ $n\pi^+$ production 
better constrain the branching ratios $\beta_{N\pi}$ and 
$\beta_{N\eta}$ for the $S_{11}(1535)$. For the first time, 
the longitudinal transition amplitude to the $S_{11}(1535)$ was extracted 
from the $n\pi^+$ data. Also, new results on the transition 
amplitudes  for the $D_{13}(1520)$ resonance are presented showing a 
rapid transition from helicity 3/2 dominance seen at the real photon point 
to helicty 1/2 dominance at higher $Q^2$. I also discuss aspects of the search 
for new excited nucleon states in strangeness photoproduction.
\end{abstract}



\begin{multicols}{2}

\section{Introduction}
\label{intro}

Electroexcitation of nucleon resonances has long been recognized as a sensitive tool
in the exploration of the complex nucleon structure at varying distances scales.  
Resonances play an important role in fully understanding the spin structure of the nucleon. 
More than 80\%  of the helicity-dependent integrated total photoabsorption cross section 
difference (GDH integral) is the result of the $N\Delta(1232)$ transition\cite{buli,mami}, 
and at a photon virtuality $Q^2=1$~GeV$^2$ more than 50\% of the first moment 
$\Gamma_1^P(Q^2) = \int_0^1{g_1(x,Q^2)dx}$ of the spin structure function $g_1$ for the 
proton is due to contributions of the resonance region at $W<2$~GeV \cite{fatemi03}, 
and is crucial for describing the entire $Q^2$ range of $\Gamma_1^p(Q^2)$
and $\Gamma_1^{p-n}(Q^2)$ for the proton and proton-neutron difference 
respectively~\cite{prok08,ioffe,deur}.  Nucleon resonances may also be responsible
for significant portion of the orbital angular momentum contributions to the nucleon spin due
to the strong excitation of resonances with quark orbital angular momentum $\Delta L_z =1$ 
and $\Delta L_z=2$ in the 2nd and 3rd resonance region.

Studies of nucleon resonances are of interest in their own rights. Studying the 
systematics of the excitation spectrum is key to understanding the symmetry underlying 
the nucleon matter.  Electroexcitation 
of resonances allows us to probe the internal structure of the excited state 
knowing the structure of the ground state. The most comprehensive predictions 
of the resonance excitation spectrum come from the various implementation 
of the symmetric constituent quark model based on broken $SU(6)$ 
symmetry~\cite{isgur}. Other models predict a different excitation spectrum, 
e.g. through a diquark-quark picture, or through 
dynamical baryon-meson interactions. The different resonance models not only 
predict different excitation spectra but also different $Q^2$ dependence of 
transition form factors.  Mapping out the transition form factors will tell 
us a great deal about the underlying quark or hadronic structure.    

The CLAS detector is the first full acceptance instrument with sufficient resolution to 
measure exclusive 
electroproduction of mesons with the goal of studying the excitation of nucleon resonances in 
detail. With the  GeV electron beam energy available at JLab, the entire resonance mass 
region and a wide range in the photon virtuality $Q^2$ can be 
studied, and many meson final states are measured simultaneously\cite{burkert-lee}. 
In this talk I discuss recent 
results from the electroproduction of single pions to determine the transition amplitudes 
to  several well-known states.  I will also discuss selected aspects of the search for new
excited baryon states.

\section{The $N\Delta(1232)$ transition}
\label{sec:ndelta}
An interesting aspect of nucleon structure at low energies 
is a possible quadrupole deformation of the nucleon's lowest excited state, the 
$\Delta(1232)$. 
Such a deformation would be evident in non-zero values of the quadrupole transition 
amplitude $E_{1+}$ from the nucleon to the $\Delta(1232)$\cite{buchmann}. 
In models with $SU(6)$ spherical symmetry, the $N\Delta$ transition is simply due to
a magnetic dipole $M_{1+}$ mediated by a spin flip, and $E_{1+} = S_{1+} = 0$ . 
Dynamically, quadrupole deformations may arise through the interaction of the photon with the pion 
cloud\cite{sato,yang} or through the one-gluon exchange mechanism\cite{koniuk}. 
Simultaneous description of 
both $R_{EM}$ and $R_{SM}$ is achieved with dynamical models that include pion-nucleon
interactions explicitly, supporting the idea that most of the 
quadrupole strength in the $N\Delta(1232)$ transition is due to meson effects. 
At the real photon point over 30\% of the transition strength is due to pion effects. 

\begin{center}
\includegraphics[width=7cm]{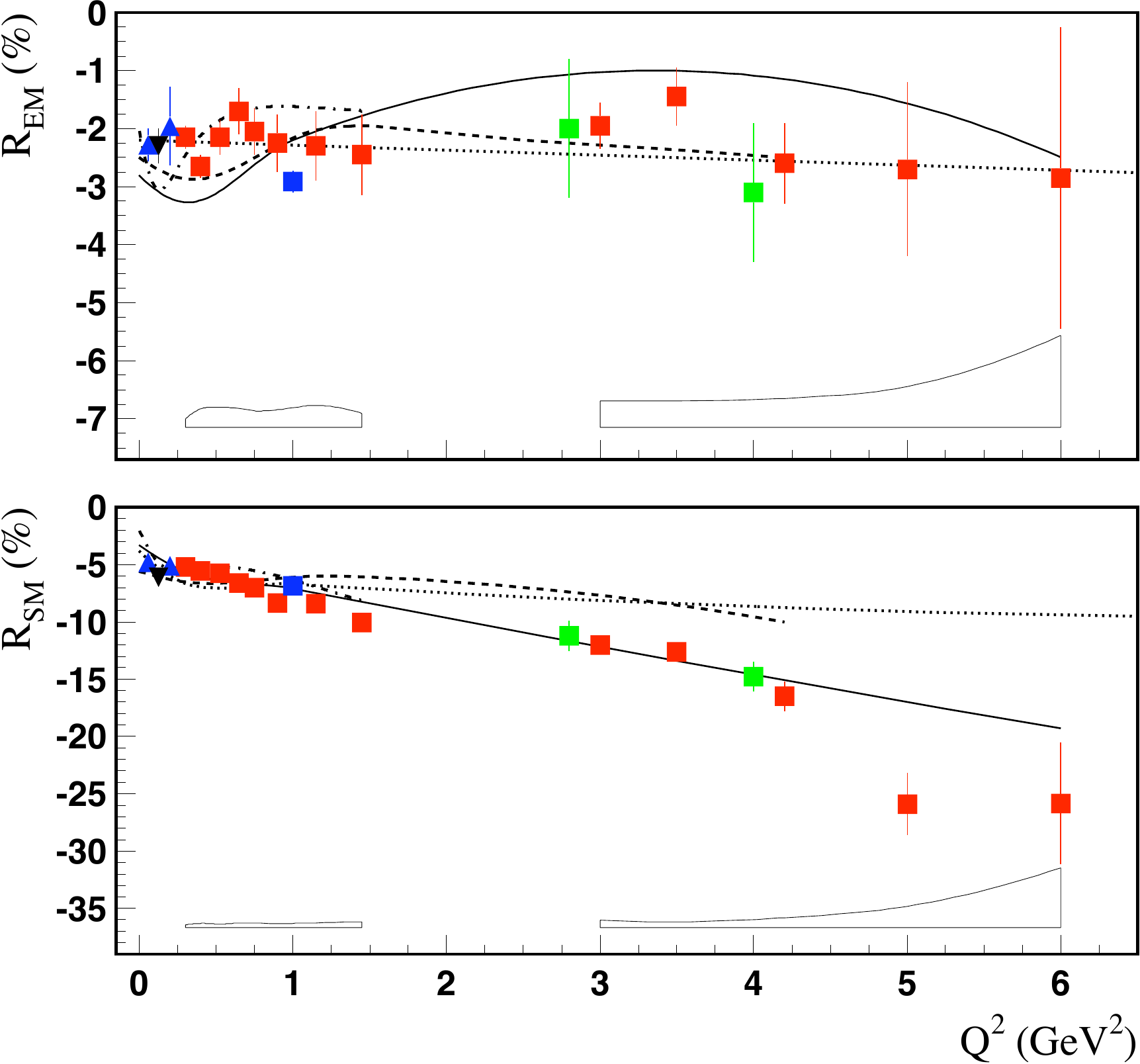}
\figcaption{\label{remrsm}$R_{EM}$ and $R_{SM}$ extracted from exclusive reactions
 $p(e,e^\prime p)\pi^0$ using unitary isobar and fixed-t dispersion relations as analysis tools.  }
 \label{remrsm}
 \end{center}
At asymptotic momentum transfer, a model-independent prediction of helicity conservation 
requires $R_{EM}\equiv E_{1+}/M_{1+} \rightarrow +1$. An interpretation of $R_{EM}$ in 
terms of a quadrupole deformation can thus only be valid at low momentum transfer.  
Results of the multipole analysis of the JLab data\cite{kjoo,frolov,ungaro,kelly} as well 
as low $Q^2$ data from MAMI\cite{mami-delta,beck}, Bates\cite{bates-delta} and LEGS\cite{legs}
are shown in Fig.\ref{remrsm}.  
A consistent picture emerges from these precise data: (1) The magnetic transition form factor drops much 
faster with $Q^2$ than the elastic 
form factor $G^p_m$ which follows approximately the dipole form $G_D$. (2) $R_{EM}$ remains negative, small 
and nearly constant in the entire range $0< Q^2<6$~GeV$^2$. (3) There are no indications that leading 
pQCD contributions are important as they would result in $R_{EM} \rightarrow +1$ \cite{carlson}, and (4) $R_{SM}$ 
remains negative, and its magnitude rises with $Q^2$.

 Ultimately, we want to come to a QCD description of these important 
nucleon structure quantities. In recent years significant effort has been  
extended towards a Lattice QCD description of the $N\Delta$ 
transition\cite{alexandrou1,alexandrou2}. 
Within the still large error bars, both quenched 
and unquenched calculations at
$Q^2<1.5$~GeV$^2$ with pion masses of 400~MeV are consistent with a small
negative value of $R_{EM}$ in agreement with the data.  For $R_{SM}$ 
there is a discrepancy at low $Q^2$ in both quenched and unquenched QCD 
calculation while the rise in magnitude of $R_{SM}$ with $Q^2$ observed in the data 
is quantitatively reproduced in full QCD at $Q^2>1$GeV$^2$.    
\begin{center}
\includegraphics[width=6cm, height=4cm]{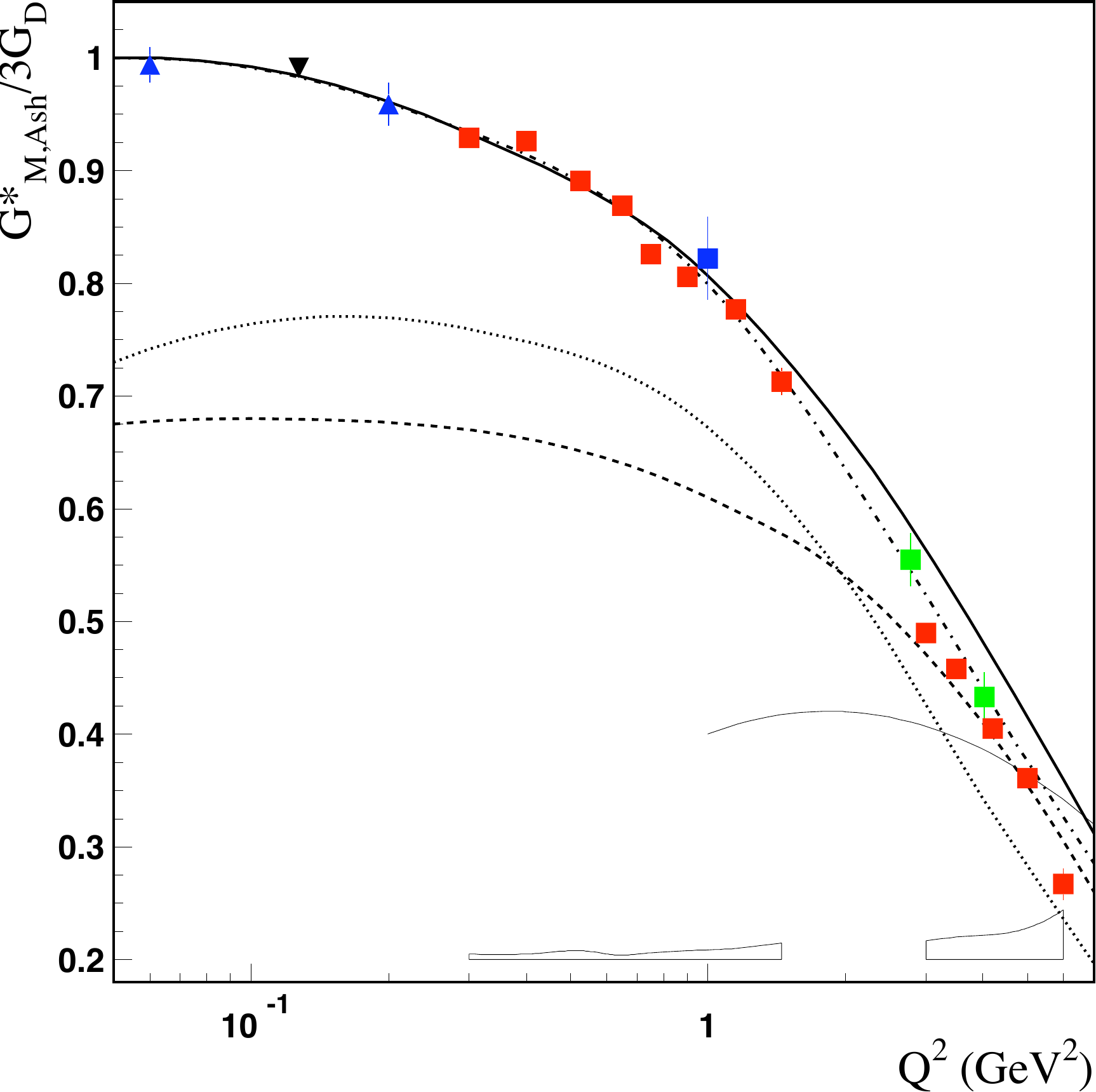}
\figcaption{Magnetic transition form factor $G_M^\Delta$ extracted from exclusive
reactions pion production, normalized to the dipole form.}
\label{gm}
\end{center}

\section{The Roper resonance $P_{11}(1440)$ - a puzzle resolved }
The $N(1440)P_{11}$ resonance (aka "Roper") has been a focus of attention for 
the last decade, largely due
to the inability of the standard constituent quark model to describe basic
features such as the mass, photocouplings, and $Q^2$ evolution. 
This has led to alternate approaches where the state is treated as 
a gluonic excitation of the nucleon\cite{libuli}, or has a small quark core 
with a large meson cloud\cite{cano}, or is a hadronic molecule of a nucleon and a 
$\sigma$ meson\cite{krewald}. 

Given these different theoretical concept for the structure of the state, 
the question ``what is the nature of the Roper state?'' has been a focus of 
the $N^*$ program with CLAS.  The state couples to both $N\pi$ and $N\pi\pi$ 
final states. It is also a very wide resonance with about 350 MeV total width. 
Therefore single and double pion electroproduction data covering a large range in the 
invariant mass W, with full center-of-mass angular coverage are key in extracting the 
transition form factors in a large range of $Q^2$. As an isospin  $I = {1\over 2}$ state, 
the $P_{11}(1440)$ couples more strongly to n$\pi^+$ than to p$\pi^o$. Also
contributions of the high energy tail of the $\Delta(1232)$ are much reduced in that 
channel due to the $I = {3\over 2}$ nature of the $\Delta(1232)$. 
\begin{center}
\includegraphics[width=8.5cm]{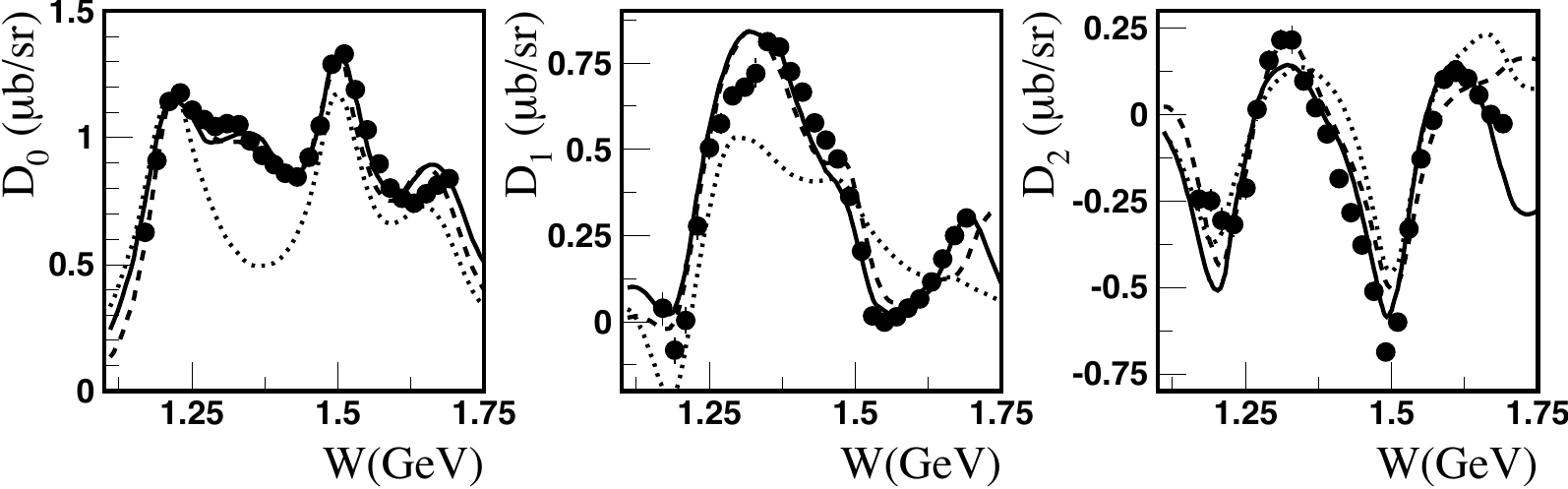} 
\figcaption{\protect W dependence of the three lowest Legendre moments from $n\pi^+$ 
angular distributions at fixed $Q^2=2.05$~GeV$^2$. The dotted line indicates the cross section
when the amplitudes of the $P_{11}(1440)$ are set equal 0.}
\label{legendre}
\end{center}
A large sample of differential cross sections and polarized beam asymmetry data from  
CLAS\cite{park08} have
been analyzed using the fixed-t dispersion relations approach (DR) and the unitary isobar 
model (UIM) \cite{janr}. Some of the features of the data may best be seen in the Legendre moments 
determined in fits to Response functions can be expressed in terms of Legendre polynomials, e.g.
the azimuthal angle independent part of the differential cross section can be written as:
$$\sigma_T + \epsilon \sigma_L = \sum_{\ell=0}^\infty D_\ell^{T+L}P_\ell({cos\Theta^*_\pi}).$$  
\noindent 
Figure~\ref{legendre} shows the lowest Legendre moments for this response functions. 
The transverse and longitudinal electro-coupling amplitudes $A_{1/2}$ and $S_{1/2}$ of 
the transition to the $N(1440)P_{11}$ resonance are extracted from fits to the 
data\cite{aznauryan-2}. They are shown in Fig.~\ref{roper}.      
\begin{center}
\includegraphics[width=8cm]{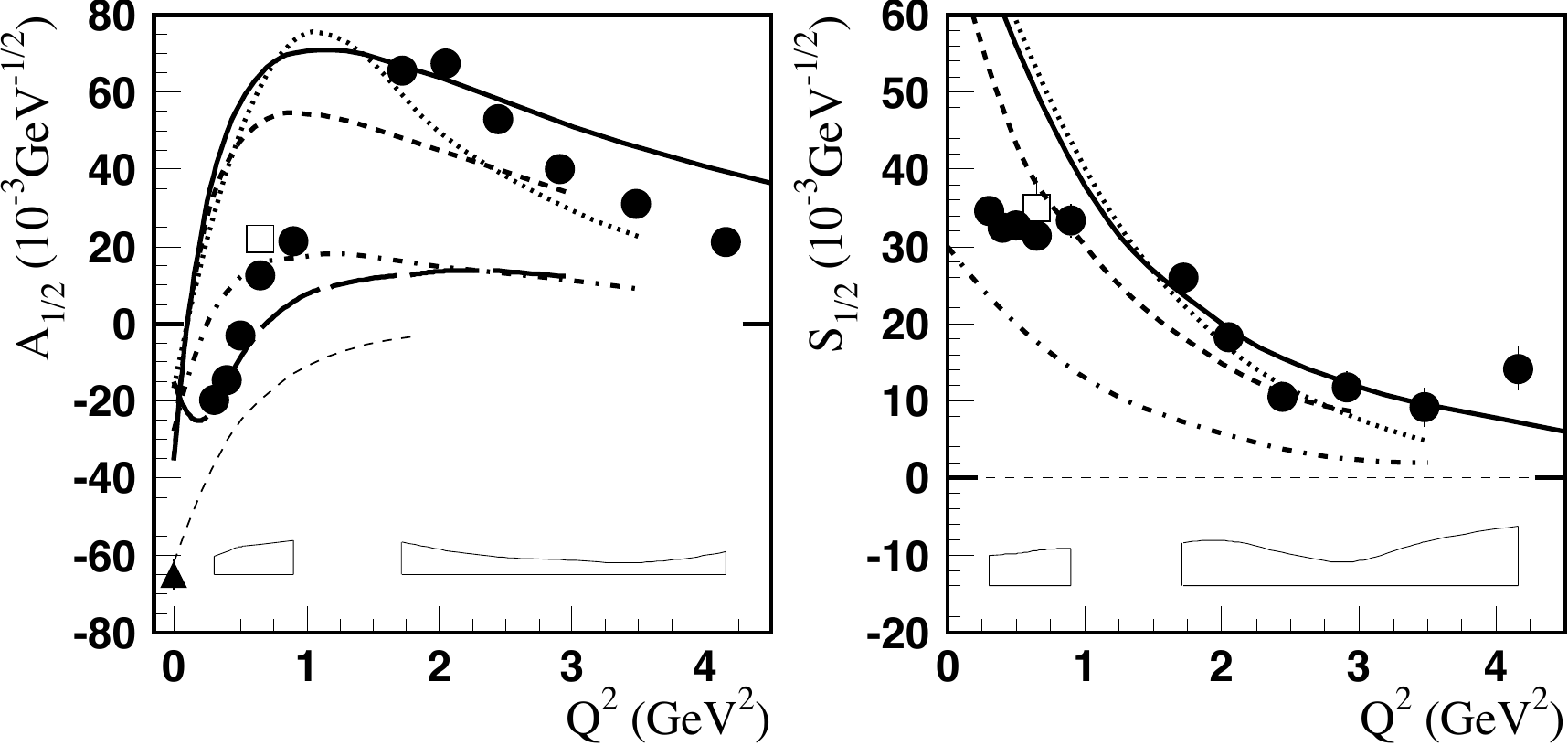}
\figcaption {Transverse electrocoupling amplitude for the 
Roper $P_{11}(1440)$ (left panel). The full circles 
are the new CLAS results. The squares are previously published results of fits to CLAS data
at low $Q^2$. The right panel shows the longitudinal amplitude. The bold curves are all 
relativistic light front quark model calculations \protect\cite{aznauryan-qm}. The thin solid 
line is a non-relativistic quark model with a vector meson cloud\protect\cite{cano}, and 
the thin dashed line is for a gluonic excitation\protect\cite{libuli}. } 
\label{roper}
\end{center}
At the real photon point $A_{1/2}$ is negative. The CLAS results show a fast rise of the 
amplitude with $Q^2$ and a sign change near $Q^2=0.5$~GeV$^2$. At $Q^2=2$GeV$^2$ the 
amplitude has about the same magnitude but opposite
sign as at $Q^2=0$. It slowly falls off at high $Q^2$. This remarkable behavior of a sign 
change with $Q^2$ has not been seen before for any nucleon. The longitudinal coupling 
$S_{1/2}$ is smaller than the transverse one. 
The first results for the transition form factors of the Roper have recently been obtained 
in unquenched QCD\cite{roper-lqcd}.
The hybrid baryon model is clearly ruled out for both amplitudes. At high $Q^2$ both 
amplitudes are qualitatively described by the 
light front quark models, which strongly suggests that the Roper is indeed a radial excitation 
of the nucleon. The low $Q^2$ behavior is not well described by the LF models and they
fall short of describing the amplitude at the photon point. This indicates that important
contributions, e.g. meson-baryon interactions at large distances may be missing. 
In the light front system the Dirac and Pauli form factors F1 and F2 can be interpreted as 
two-dimensional charge densities ~\cite{vdh08}. Figure~\ref{p11_densities} shows the transition charge density in transverse impact parameter space.
\begin{center}
\includegraphics[width=5cm]{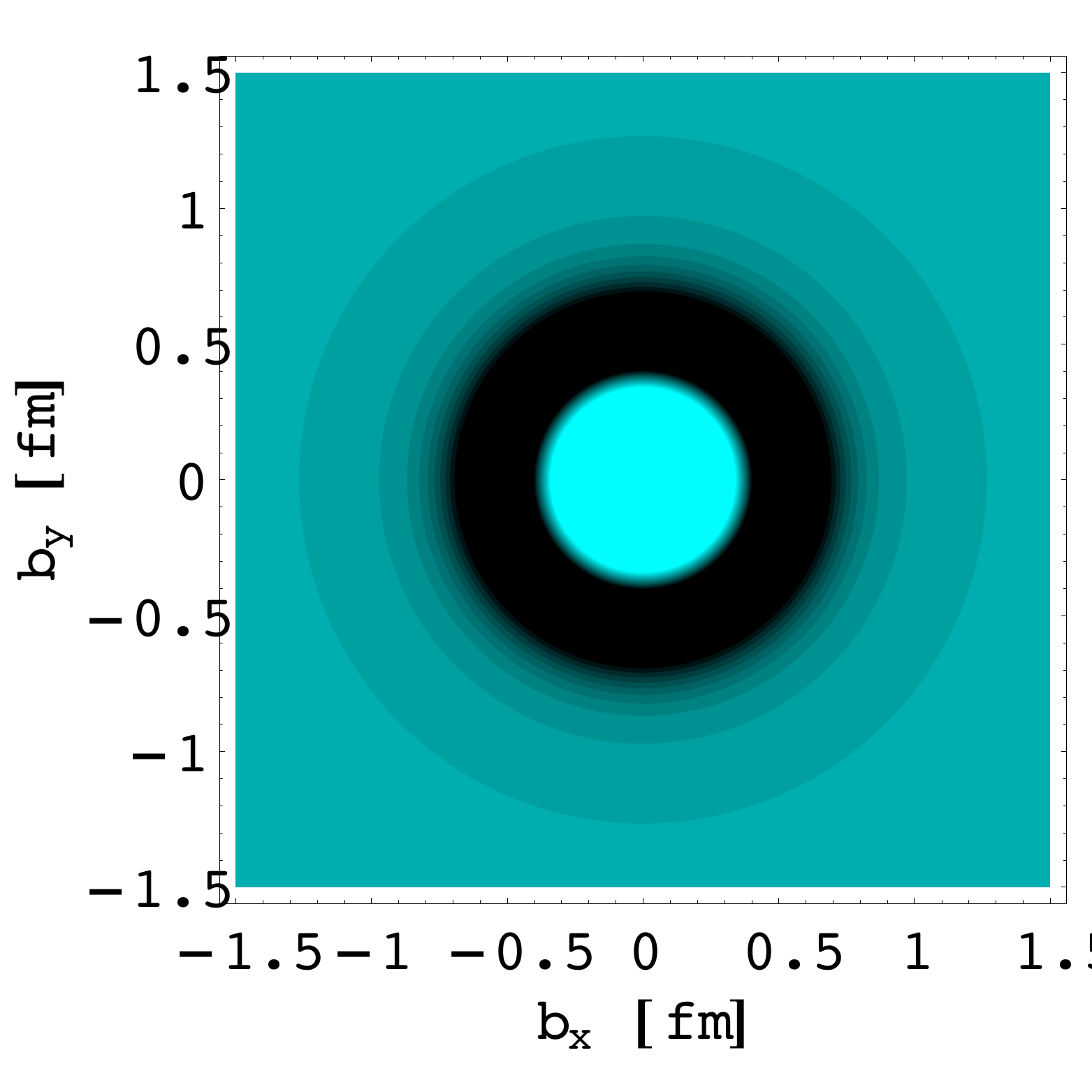}
\figcaption {Transverse transition charge densities in the light front system for the proton to Roper
transition in the light-front helicity $+1/2 \rightarrow +1/2$. The light area indicates dominance of 
up quarks (positive charge), the dark area is due to down quarks (negative charge). } 
\label{p11_densities}
\end{center}

\section{The $N(1535)S_{11}$ state} 
This state has been studied in the $p\eta$ channel where it 
appears as a rather isolated resonance near the $N\eta$ threshold.
Data from CLAS\cite{thompson,denizli} and Hall C\cite{armstrong} 
have provided a consistent picture of the $Q^2$ evolution obtained 
from $\eta$ electroproduction data alone.  
There are two remaining significant uncertainties that need to be examined. 
The first one is due to the 
branching ratio of the $S_{11}(1535) \rightarrow p\eta$, the second one is due 
to the lack of precise information on the longitudinal coupling, which in the $p\eta$ channel
is usually neglected. 

The PDG gives a range of  $\beta^{PDG}_{N\eta} = 0.45 - 0.60$. 
Since this state practically does not couple to channels other than $N\eta$ and $N\pi$, 
a measurement of the reaction $ep\rightarrow e\pi^+n$ will reduce this uncertainty.
Also, the $N\pi$ final state is much more sensitive to the longitudinal amplitude due to
a strong $S_{11}-P_{11}$ interference term present in the $N\pi$ channel. 
Adjusting $\beta_{N\pi} = 0.45$ and $\beta_{N\eta} = 0.45$ brings the two data sets into 
excellent agreement as shown in Fig.~\ref{s11}. 
\begin{center}
\includegraphics[width=8cm]{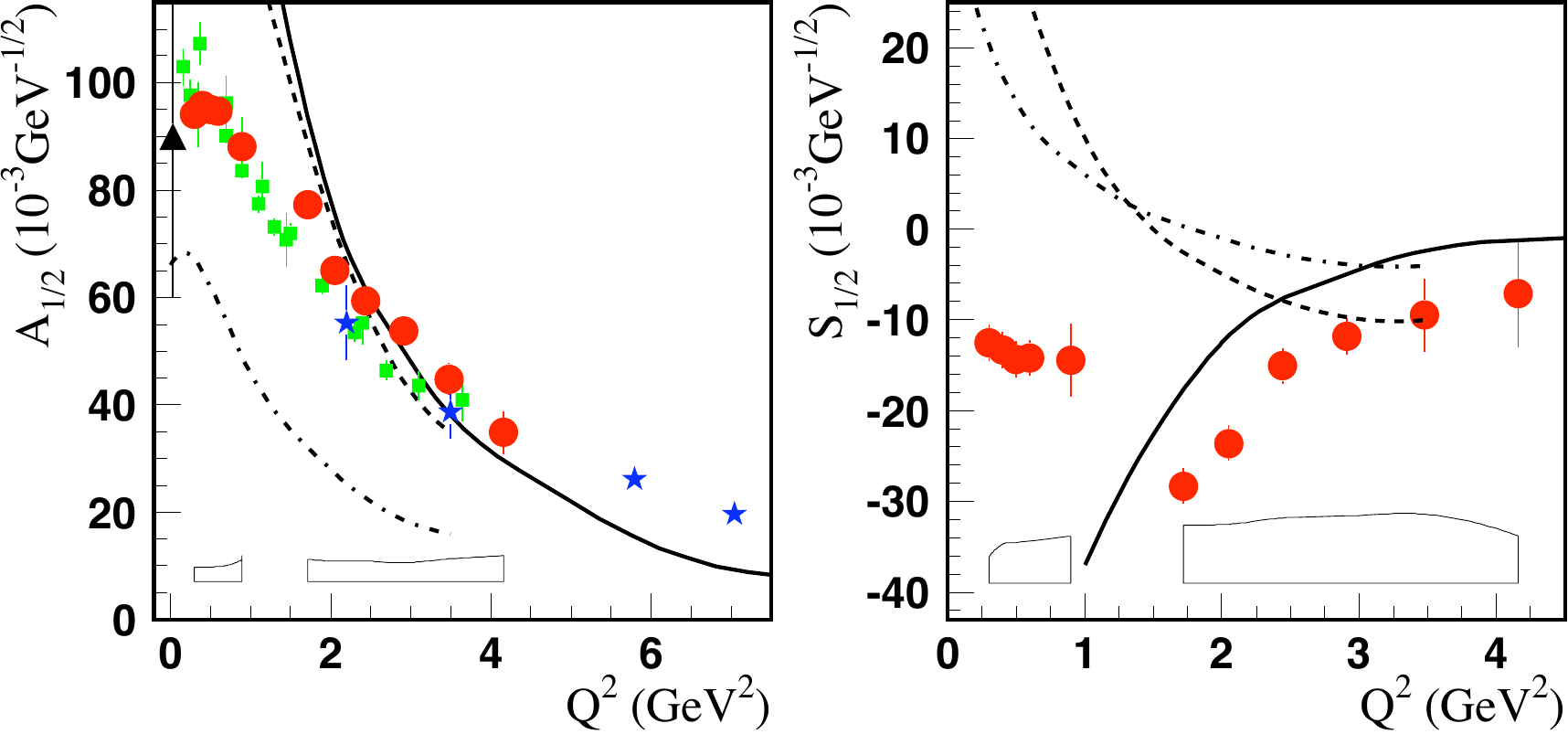} 
\figcaption {The transition amplitude $A_{1/2}$ for the $S_{11}(1535)$. The big full circles are 
from the analysis of the CLAS $n\pi^+$ and $p\pi^\circ$ data\protect\cite{aznauryan-1,aznauryan-2}. 
The other data are from the analysis of $p\eta$ data~\protect\cite{thompson,armstrong,denizli,dalton}. 
The theory curves are from various constituent quark models\protect\cite{capstick,salme,merten,santopinto,warns}.}
\label{s11}
\end{center}
An important advantage of the $N\pi$ channel is that it is also 
sensitive to the longitudinal transition amplitude which is due to a significant $s-p$ wave interference 
with the nearby $p$-wave amplitude of the $P_{11}(1440)$. This can be seen in the multipole 
expansion of the Legendre moment $D_0^{LT}$:
 $$D_0^{LT} = {{|\vec{q}|} \over K} Re(E_{0+}S_{1-}^* + S_{0+}M_{1-}^*).$$   
 
\section{Helicity structure of the $D_{13}(1520)$ }

A longstanding prediction~\cite{close} of the dynamical constituent quark model is the rapid helicity 
switch from the dominance of the $A_{3/2}$ at the real photon point to the dominance of 
the $A_{1/2}$ amplitude at $Q^2 > 1$~GeV$^2$. In the simple non-relativistic harmonic 
oscillator model with spin and orbit flip amplitudes only, the ratio of the two amplitudes 
is given by: 
$$ {A_{1/2}^{D13} \over A_{3/2}^{D13}} = {-1 \over \sqrt{3}}({\vec{Q}^2 \over \alpha} - 1)~,$$ 
where $\alpha$ is a constant adjusted to reproduce the ratio at the photon point where $A_{1/2}$ is 
very small.
\begin{center}
\includegraphics[width=8.5cm, height=3.5cm]{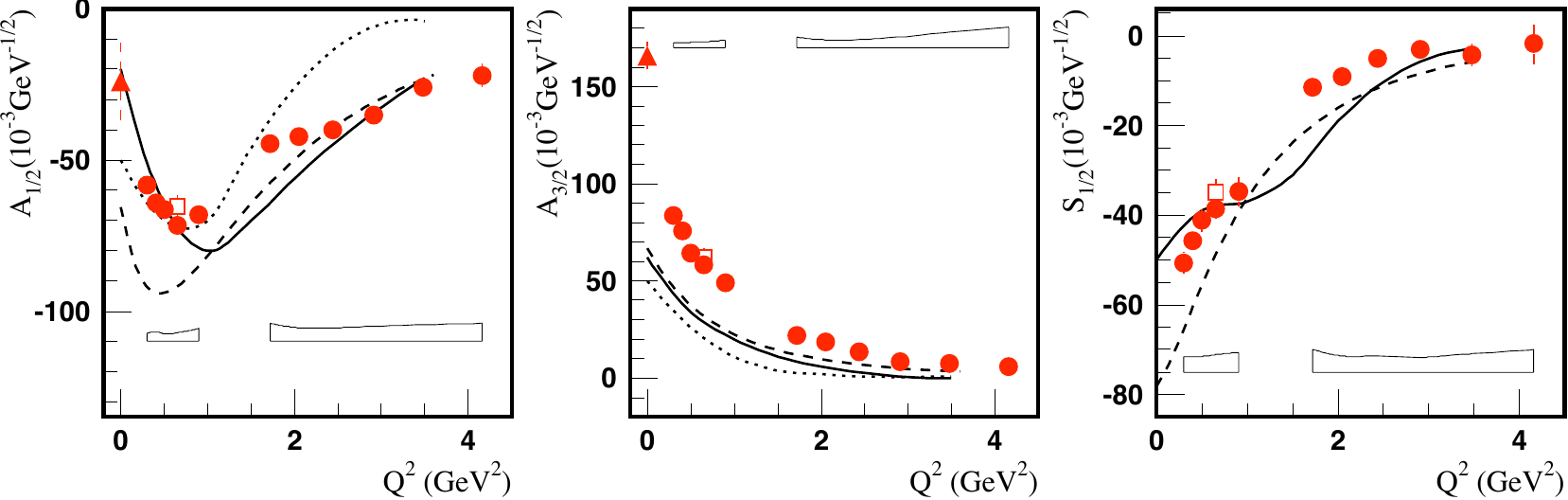}
\figcaption{Electro-coupling amplitudes $A_{1/2}$ (left) and $A_{3/2}$ 
(middle) and $S_{1/2}$ (right) for the $D_{13}(1520)$. Model curves as in Fig.6.} 
\label{d13}
\end{center} 
Figure~\ref{d13}
shows the CLAS results for the electro-couplings. The $A_{3/2}$ amplitude is large at the real 
photon point and decreasing rapidly 
in strength with increasing $Q^2$. The $A_{1/2}$ amplitude increases rapidly in magnitude 
with increasing $Q^2$, before falling off slowly at $Q^2>1$~GeV$^2$. The rapid switch in the 
helicity structure can best be seen in the helicity asymmetry defined as 
$A_{hel}= { (A_{1/2}^2 - A_{3/2}^2) / (A_{1/2}^2 + A_{3/2}^2)}$, and shown in Fig.~\ref{hel_asym}.
\begin{center}
\includegraphics[width=4cm]{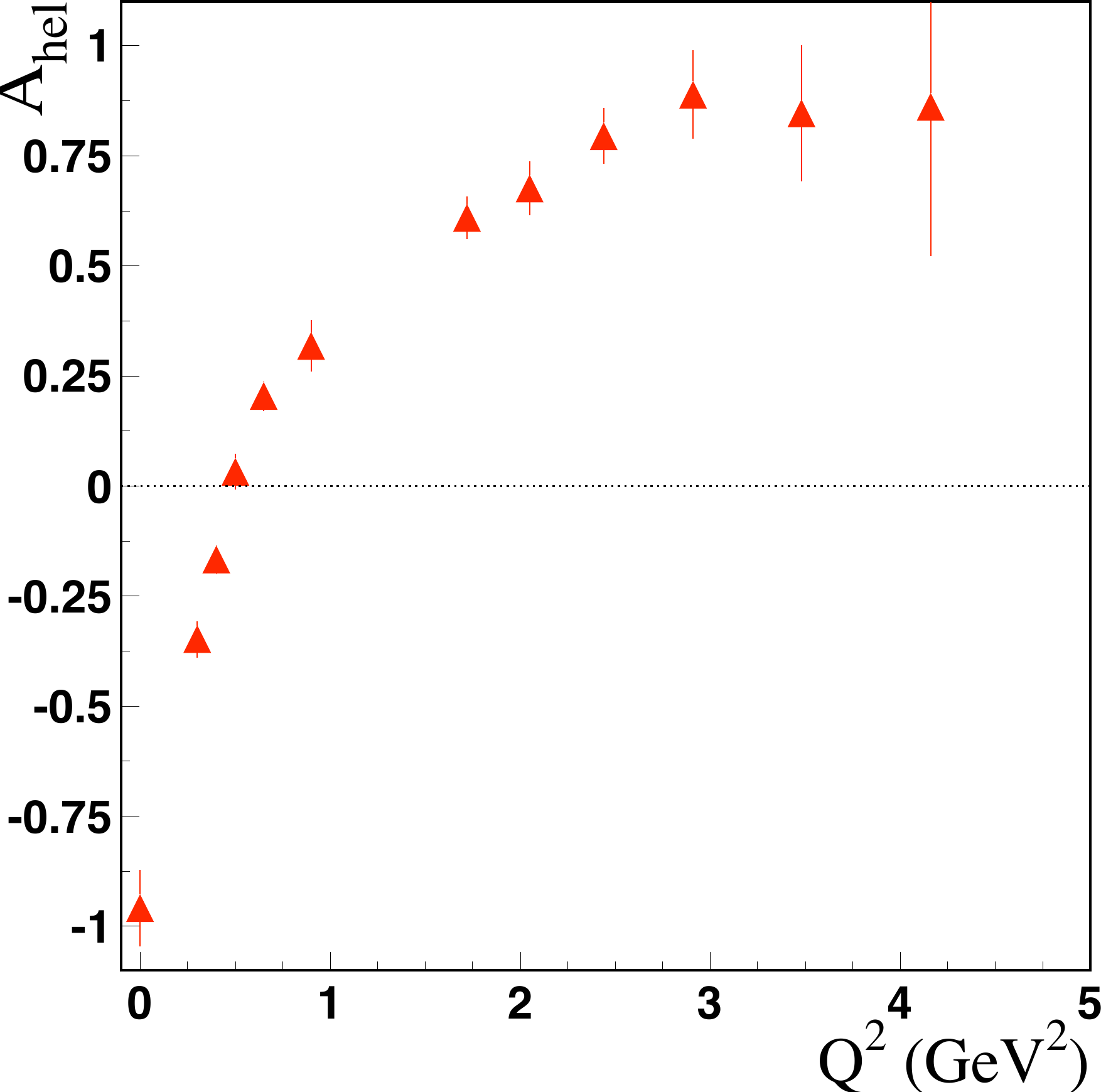}
\figcaption{Helicity asymmetry for the photon induced proton-$D_{13}$ transition vs $Q^2$.} 
\label{hel_asym}
\end{center}

\section{Search for new nucleon states}    
The experimental program makes use of the CEBAF Large Acceptance Spectrometer (CLAS)
~\cite{clas} which provides particle identification and
momentum analysis in a polar angle range from 8$^\circ$ to 140$^\circ$. The 
photon energy tagger provides energy-marked photons with an energy resolution of 
${\sigma(E) \over E} = 10^{-3}$. Other equipment includes a coherent 
bremsstrahlung facility with a goniometer for diamond crystal positioning and 
angle control. The facility has been used for linearly polarized photons with 
polarizations up to 90\%. There are two frozen spin polarized 
targets, one based on butanol as target material (FROST), 
and one using HD as a target material (HD-Ice). The latter is currently under construction.
FROST has already been operated successfully in 
longitudinal polarization mode and will be used in transverse polarization 
mode in 2010.
The HD-Ice target will be used as polarized neutron (deuteron) target in scheduled runs in 2010 and
2011. Circularly polarized photons can be generated by scattering the highly polarized electron beam from 
an amorphous radiator.

It is well understood that measurements of differential cross sections in 
photoproduction of single pseudoscalar mesons alone results in ambiguous solutions 
for the contributing resonant partial waves. The $N^*$ program at JLab is therefore 
aimed at complete, or nearly complete measurements for processes 
$\vec{\gamma} \vec{p} \rightarrow \pi N, ~\eta p,~K^+\vec{Y}$ and 
$\vec{\gamma} \vec{n} \rightarrow \pi N, ~K\vec{Y}$. Complete information can be 
obtained by using a combination of linearly and circularly polarized photon beams, 
measurement of hyperon recoil polarization, and the use targets with longitudinal 
and transverse polarization. The reaction is fully described by 4 complex
 parity conserving amplitudes, requiring 8 combinations of 
beam, target, and recoil polarization measurements for an 
unambiguous extraction of the production amplitudes. If all possible combinations are 
measured, 16 observables can be extracted. In measurements that involve nucleons 
in the final state where the recoil polarization is not measured, 7 independent observables 
can be obtained directly, and the recoil polarization asymmetry $P$ can be inferred 
from the double polarization asymmetry with linearly polarized beam and transverse target 
polarization. 
\begin{center}
\includegraphics[width=8cm]{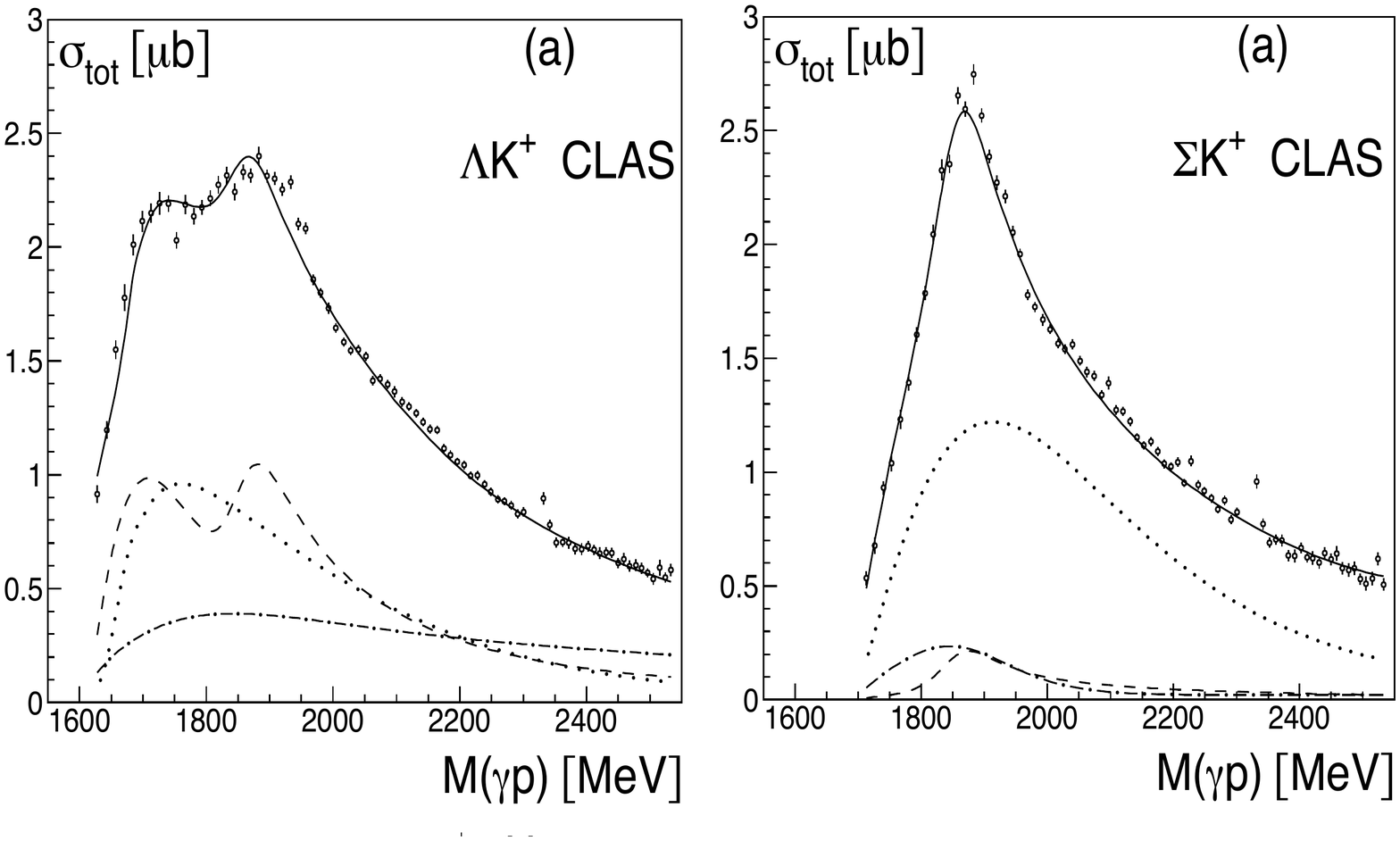}
\figcaption{Integrated total cross section data for $K^+\Lambda$ and $K^+\Sigma^\circ$ channels. 
The lines represent the fit results of the Bonn-Gatchina coupled-channel analysis. The dashed line shows
the energy-dependence of the $P_{13}$ partial wave which indicates the presence of two $P_{13}$ resonances 
at 1720 MeV and at 1900 MeV, respectively.  In this analysis, the new $P_{13}(1900)$ contributes a significant 
fraction to the total $K^+\Lambda$ cross section.}
\label{fig:lambda_int_cs}
\end{center} 

\subsection{Search for new states in strangeness channels} 
A large amount of cross section data have been collected in recent years on the $K\Lambda$ and 
$K\Sigma$ photo-production~\cite{mcnabb04,brad06}. These data cover the nucleon resonance region
in fine steps of about $10$~MeV in the hadronic mass W, and nearly the entire polar angle range. 
The integrated cross section shown in 
Fig.~\ref{fig:lambda_int_cs} reveals a strong bump around $W=1900$~MeV which may be the result of s-channel
resonance production. However, more definite conclusions can only be drawn when more polarization observables
are included in the analysis.  
Preliminary results are available from the beam asymmetry in $\Lambda$ and 
$\Sigma$ production with a linearly polarized photon beam. A sample of the data is shown in 
Fig.~\ref{fig:g8b_asymmetry}.  Nearly full angle coverage with excellent statistics has been achieved. 
Beam asymmetry data are particularly important for identifying spin-parity assignment
of contributing resonances. These data will span the resonance mass region from threshold up to $W=2.3$~GeV.  
\begin{center}
\includegraphics[width=6cm]{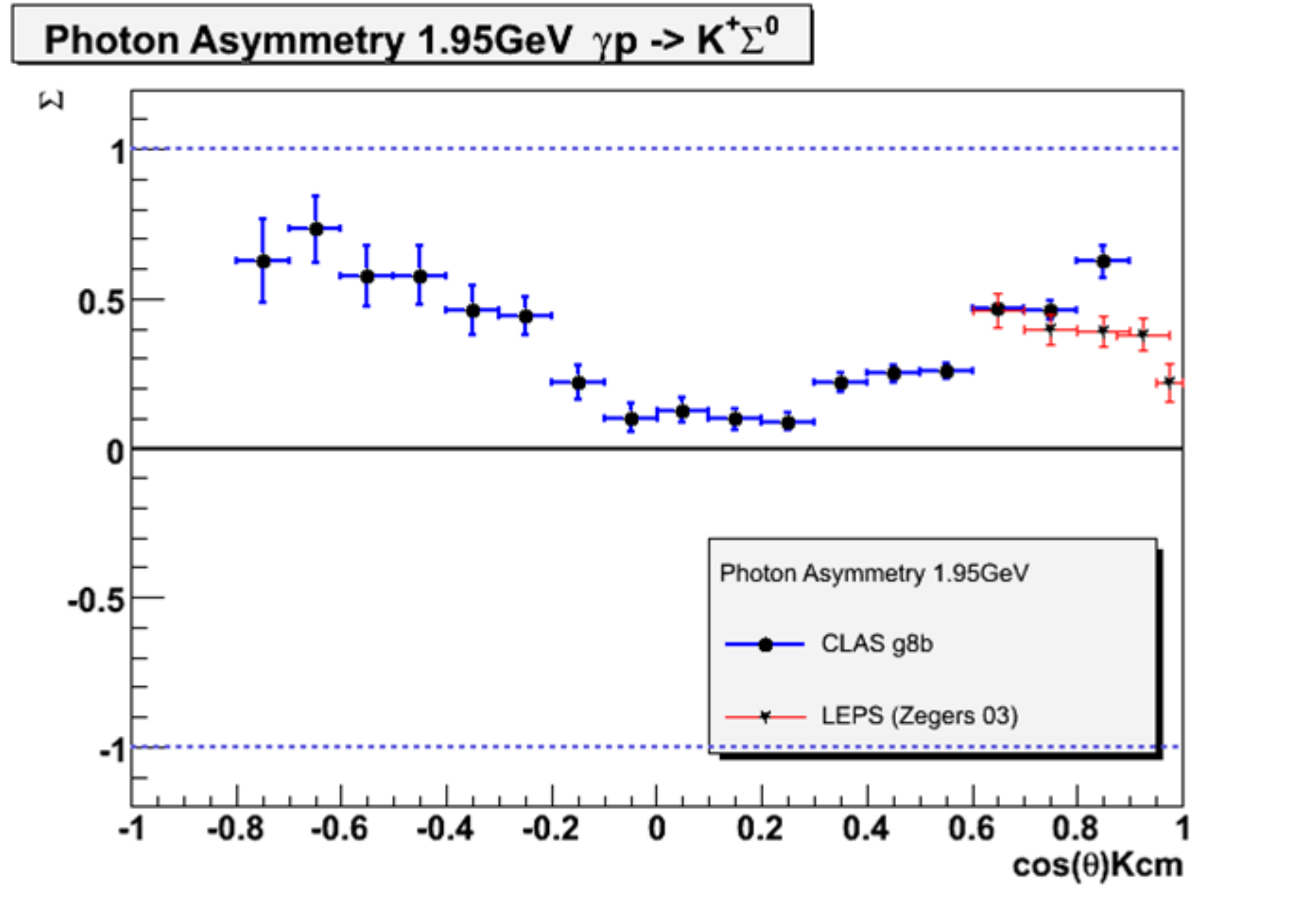}
\figcaption{Sample of preliminary CLAS data for the beam asymmetry of the $K\Sigma^\circ$ final state.}
\label{fig:g8b_asymmetry}
\end{center}  
In addition to precise $K\Lambda$ and $K\Sigma$
cross section data recoil polarization and polarization transfer data have been 
measured~\cite{brad07}. The recoil polarization data in the $K^+\Lambda$ sector showed a 
highly unexpected behavior: The spin transfer from the circularly polarized photon to the $\Lambda$ 
hyperon is complete, leaving the $\Lambda$ hyperon 100\% polarized, as can be seen in Fig.~\ref{fig:lambda_spin}. 
The quadratic sum of all polarization components $R^2 \equiv P^2 + C_x^2 + C_z^2$, 
which has an upper bound of 1, 
is consistent with $R^2 = 1$ throughout the region covered by measurement. At first glance this 
result seems adverse to the idea that the $K\Lambda$ final state has a significant component of 
$N^*$ resonance associated with. However, the analysis of the combined CLAS differential 
cross section and polarization 
transfer data by the Bonn-Gatchina group~\cite{nikonov08} shows strong sensitivity to a 
$P_{13}(1900)$ candidate state. The decisive ingredient in this analysis are the CLAS spin transfer 
coefficients $C_x$ and $C_z$ . Note that the peak observed in the $K^+\Lambda$ data seen near 1900 MeV in 
Fig.~\ref{fig:lambda_int_cs} was originally attributed to a $D_{13}(1900)$
resonance before the spin transfer data were included in the fit. A $P_{13}(1900)$ is listed as a 2-star candidate
state in the 2008 edition of the RPP~\cite{pdg2008}. If this assignment is confirmed in future
analyses which should include additional polarization data, the existence of a $N(1900)P_{13}$ 
state would be strong evidence against the quark-diquark model~\cite{santo05}, but would support the 
symmetric $SU(6)$ constituent quark model.~\cite{capstick}.  

First differential cross sections of the channel $\gamma p \rightarrow K^{*\circ}\Sigma^+$, 
covering the mass range $W > 2100$~MeV have been measured recently~\cite{hleiqawi07}. 
Several other strangeness channels such as 
$\gamma n \rightarrow K^\circ\Lambda$, $K^{*\circ} \Lambda$, $K^\circ\Sigma^\circ$, $K^+\Sigma^-$, 
and $K^+\Sigma^-(1385)$ are currently being studied to search 
for new states on neutron targets using circularly and linearly polarized photon beams.  For most of 
these channels a complete kinematical reconstruction in the neutron rest frame is possible, 
thus eliminating the effect of Fermi motion in the deuteron nucleus. Recoil polarization measurements in all 
hyperon final states are also available for the analysis.
\begin{center}
\includegraphics[width=8cm,height=6cm]{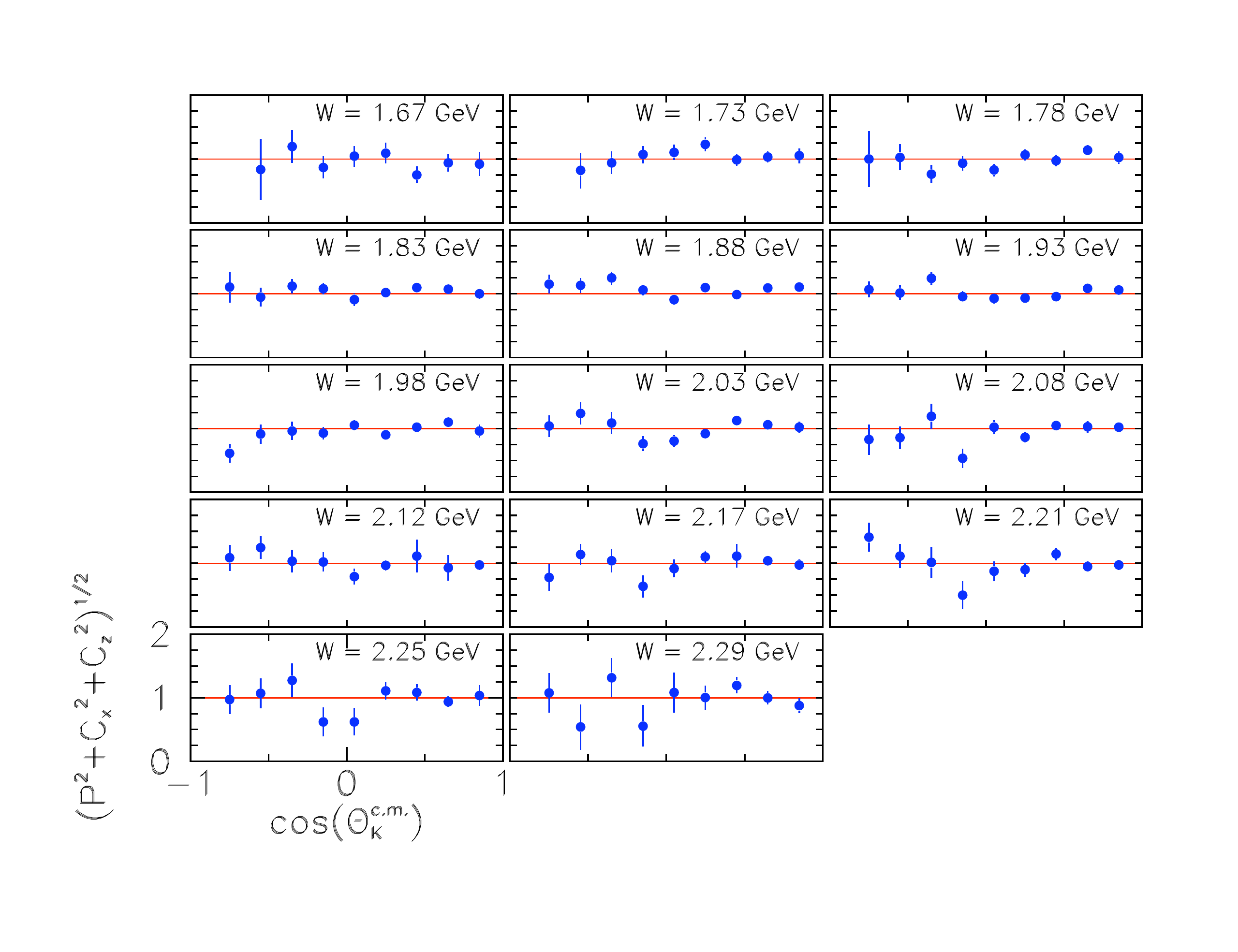}
\figcaption{\protect Spin transfer from the polarized photon to the final state $\Lambda$. The spin transfer  
data were fitted by the Bonn-Gatchina group simultaneously with the CLAS $K^+\Lambda$ and $K^+\Sigma$ differential 
cross section data. A $P_{13}(1900)$ state is required for the best fit to the cross section 
and the spin transfer data.}
\label{fig:lambda_spin}
\end{center}  

\section{Search for new cascade baryons}

The production of $\Xi$ hyperons, i.e. strangeness $S = -2$ excited states, presents another 
promising way of searching for new baryon states. The advantages of cascade hyperon states 
are due to the 
expected narrow widths of these states compared to $S = 0$, and $S = -1$ resonances. 
Possible production mechanisms include t-channel $K$ or $K^*$ 
exchanges on proton targets with an excited hyperon $Y^*$ ($\Lambda^*$ or $\Sigma^*$) as 
intermediate state and 
subsequent decays $Y^* \rightarrow K^+ \Xi^*$ and $\Xi^* \rightarrow \Xi \pi$ or $\Xi \rightarrow \Lambda (\Sigma) \bar{K}$.  
Missing mass technique may be used to search for new states in the reaction $\gamma p \rightarrow K^+ K^+ X$ if the state is 
sufficiently narrow to be observed as a peak in the missing mass spectrum. However, an analysis of the 
final state is needed to assign spin and parity to the state. Data from CLAS taken at 3.6 GeV beam energy 
show that  one can 
identify the lowest two cascade states this way.~\cite{guo07}. To identify the higher mass states higher energy is needed. 

Another approach to isolate excited cascades and determine their spin-parity, is to measure 
$\Xi$ with additional particles in the 
final state. The invariant mass of the $\Xi^\circ \pi^-$ system is displayed in Fig.~\ref{fig:cascade} and
shows the first exited state $\Xi(1530)$ and indications of additional structure near 1620 MeV.  A state 
near that mass is predicted~\cite{oset} as a dynamically generated $\Xi\pi$ system. The 
data have insufficient statistics and were taken at too low beam energy to allow further investigations. New 
data taken at 5.7 GeV electron energy and with higher statistics are currently being analyzed, and should 
allow more definite conclusions on a possible new $\Xi$ state at that mass.
\begin{center}
\includegraphics[width=8cm]{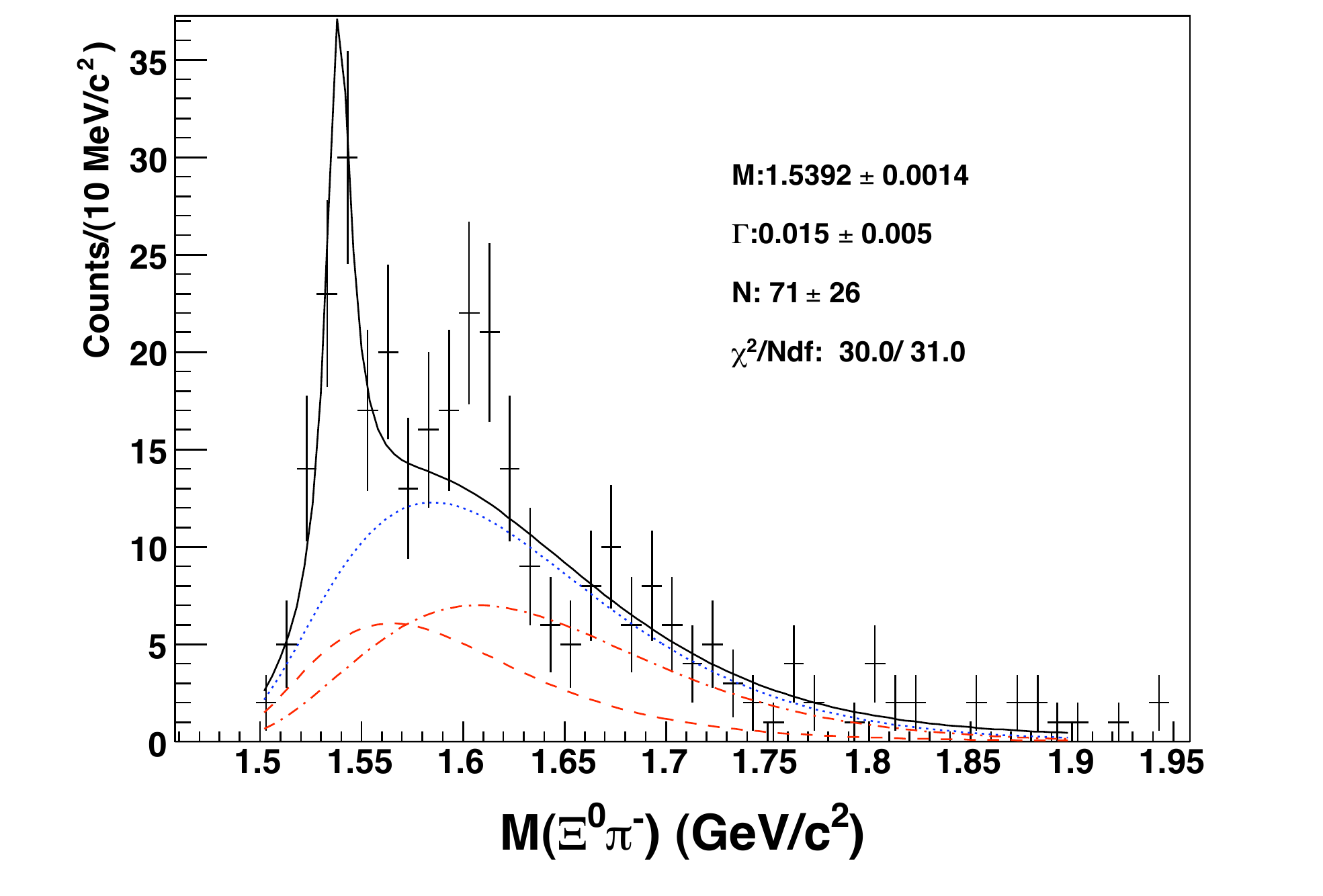}
\figcaption{Invariant mass of the $\Xi\pi^-$ final state. The $\Xi(1530)$ is clearly seen. 
A possible structure near 1620 MeV could be related to a dynamically generated $\Xi$ state~\protect\cite{oset} }
\label{fig:cascade}
\end{center} 
\section{Strangeness electroproduction}
Significant effort has been devoted to the study of electroproduction of hyperons~\cite{ambros07,carman09} as a complementary
 means of searching for new excited nucleon states. Figure~\ref{fig:Sigma0el} shows the dependence of the response functions 
on the hadronic mass W. 
\begin{center}
\includegraphics[width=8cm,height=6.5cm]{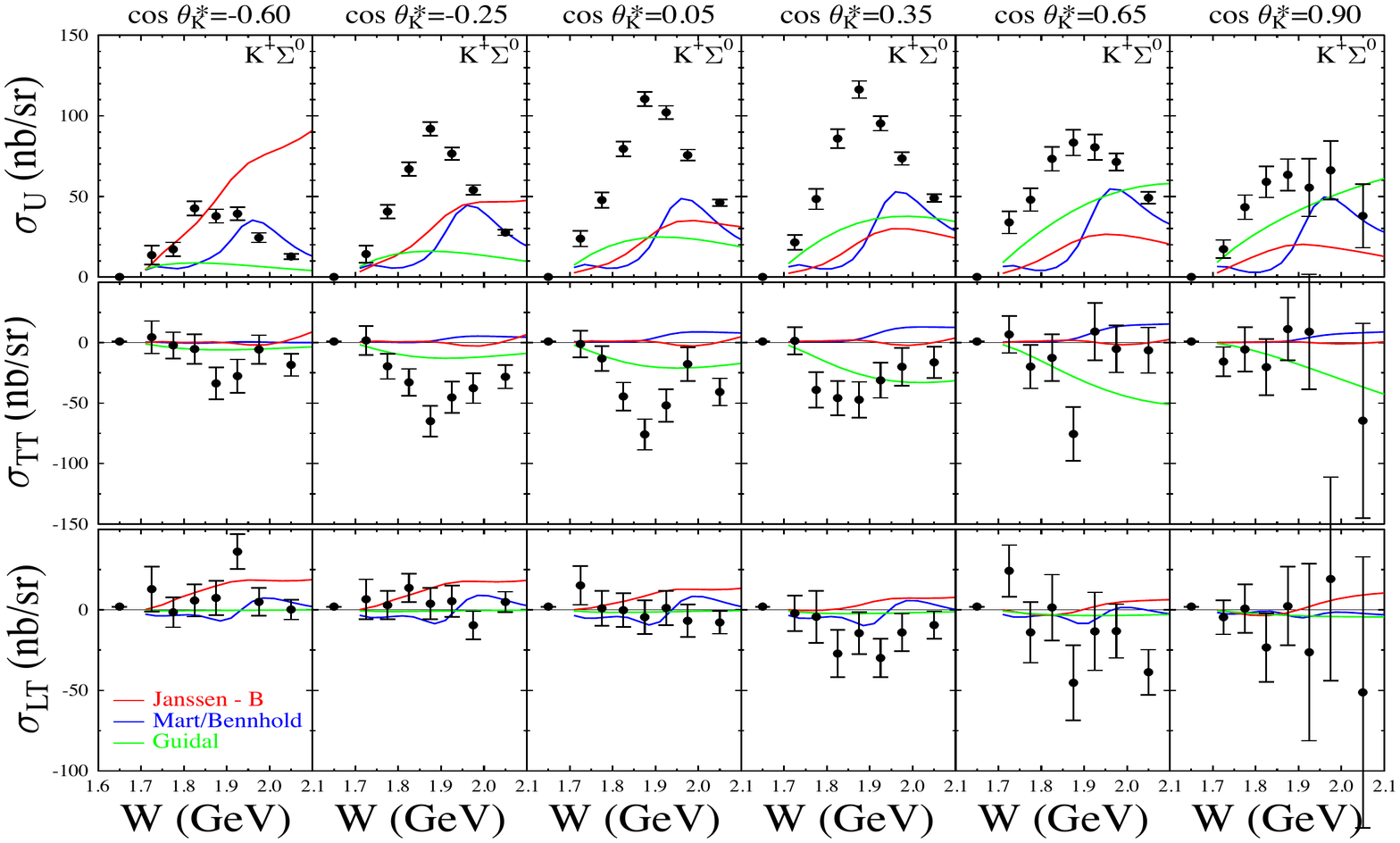}
\figcaption{\protect Separated response functions for $K^+\Sigma^\circ$ electroproduction. The response function $\sigma_u$ 
and $\sigma_{TT}$ show strong resonance-like structure near $W=1.9$GeV, which is not reproduced by any model that include 
known resonances.}
\label{fig:Sigma0el}
\end{center} 
The structure function $\sigma_u$ shows a resonance-like enhancement near 1.85 - 1.90 GeV in all
angle bins. A comparison with model calculations reveals large discrepancies. 
None of the models that include
known nucleon resonance couplings to $K\Sigma$, is able to get the normalization correct. This leaves much room for yet to be
identified resonance strength.         

\section{Outlook}
The power of the polarization measurements on the extraction of resonant  amplitudes
will be brought to bear fully when 
double and triple polarization data are available making use of longitudinally and transversely 
polarized proton and neutron targets. Data from CLAS on the polarized proton target (FROST) 
combined with linearly and circularly polarized photons are currently in the analysis stage. 
Measurements with transverse polarized proton targets are planned for 2010. 

Measurements of polarization observables on polarized neutrons are planned for 2010/2011. Some
projected data using the HD-Ice target facility in the CLAS detector are shown in 
Fig.~\ref{fig:HD_proj}. Single polarization observables $P$, $\Sigma$ and 
double polarization asymmetries for beam-recoil polarization $O_{y'}$, and  target-recoil polarization 
$T_{z'}$ are shown. Other observables, e.g. target asymmetry $T$, beam-target asymmetries $E,~F$, 
and $G,~H$ will be measured as well. The projections are for one photon energy bin out of over 25 bins. 
The solid line is the projection of the kaonMAID code~\cite{kaonmaid}.

\begin{center}
\includegraphics[width=8cm]{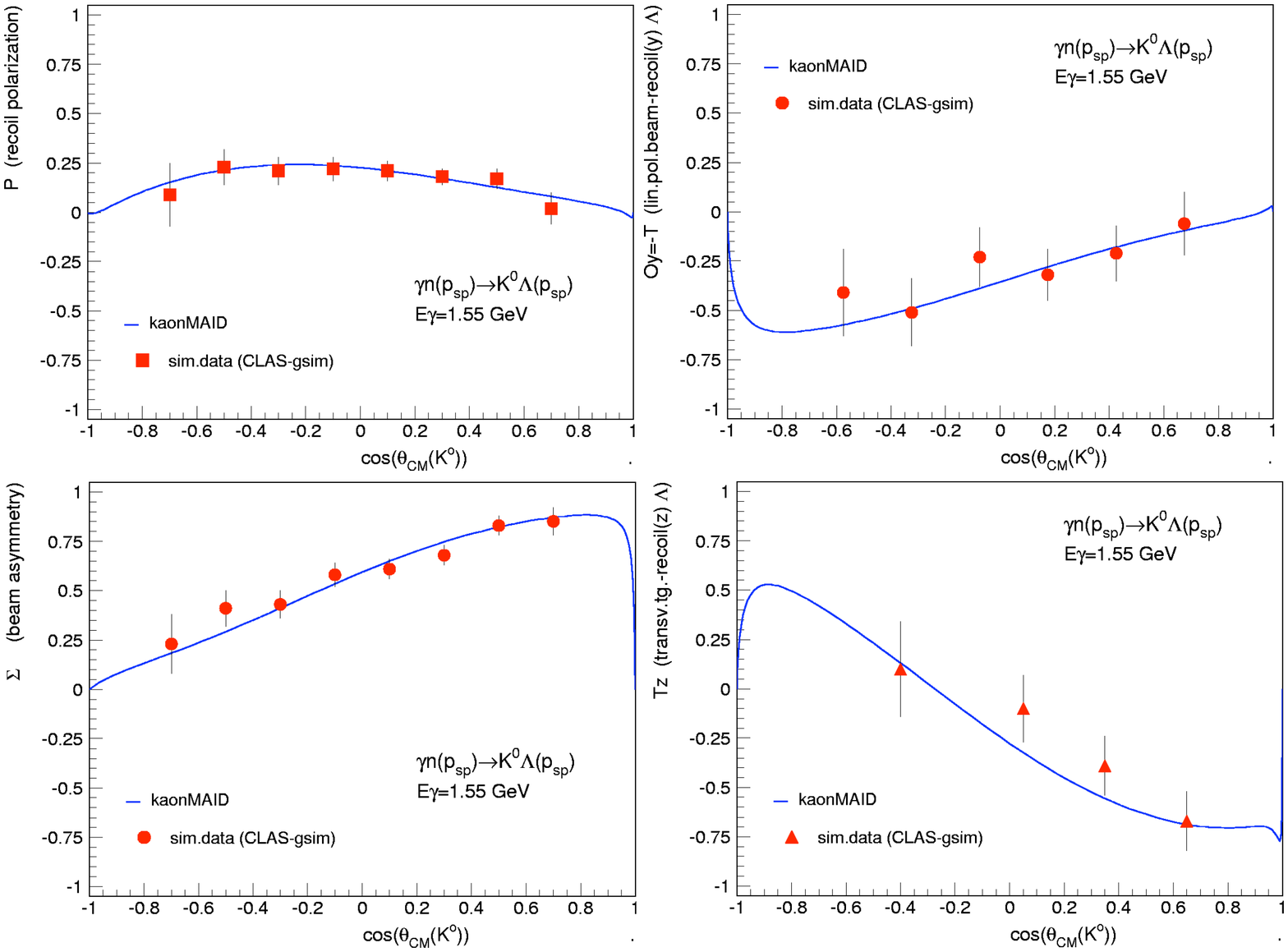}
\figcaption{Simulated data samples for the process $\gamma \vec{n} \rightarrow K_s^\circ \vec{\Lambda}$.
The uncertainties are estimated with the expected run conditions
for the upcoming polarized target run using the HD-Ice facility at JLab. }
\label{fig:HD_proj}
\end{center}  

\section{Conclusions} 
\label{sec:conclusions}
With the recent precise data on pion and eta electroproduction, combined with the 
large coverage in $Q^2$,  W, and center-of-mass angle, the study of nucleon resonance 
transitions has become an effective tool in the exploration of nucleon structure 
in the domain of strong QCD and confinement. We have learned that the $\Delta(1232)$ 
exhibits an oblate deformation. The multipole ratios $R_{EM}$ an $R_{SM}$ show no sign
of approaching the predicted asymptotic behavior, which provides a real challenge for model
builders. Lattice QCD simulations do not yet reach the high $Q^2$ covered by the 
data.  The latest data from CLAS on charged pion production reveal a sign change 
of the transverse  amplitude for the N-Roper transition near $Q^2 = 0.5$~GeV$^2$, 
and give strong evidence for this state as the first radial excitation of the nucleon. 
The hard transition form factor 
of the $S_{11}(1535)$ previously observed only in the $p\eta$ channel is confirmed in 
the $n\pi^+$ channel, which also allows us to extract the so far unmeasured longitudinal 
amplitude $S_{1/2}$. The $D_{13}(1520)$ clearly exhibits the helicity flip 
behavior predicted by the constituent quark model.  In the light front frame the 
$F_1$ and $F_2$ transition form factors that are obtained from the resonance helicity 
amplitudes, can be interpreted as 2-dimensional light cone transition charge densities. 
This development marks real 
progress in our understanding of the spatial and spin structure of the nucleon in the domain
of strong QCD.   

The search for new excited states of the nucleon has entered a new and exciting phase. 
The main goal of complete or even over determined measurements for pseudo-scalar meson 
production is now in reach with the availability of energy-tagged linearly and circularly 
polarized photon beams and the construction of polarized proton and neutron targets 
that can be used in longitudinal and in transverse polarization modes. Moreover, for processes 
involving hyperons in the final state, the hyperon recoil polarization can be measured
due to the large analyzing power of the hyperon weak decay $\Lambda \rightarrow p \pi^-$.

\acknowledgments{Authored by The Southeastern Universities Research 
Association, Inc. under U.S. DOE Contract No. DE-AC05-84ER40150 . The U.S. Government 
retains a non-exclusive, paid-up, irrevocable, world-wide license to publish or reproduce 
this manuscript for U.S. Government purposes.}

\end{multicols}

\vspace{-2mm}
\centerline{\rule{80mm}{0.1pt}}
\vspace{2mm}

\begin{multicols}{2}

\end{multicols}
\end{document}